\documentclass[a4paper,11pt]{article}
\usepackage[english]{babel}
\usepackage[utf8]{inputenc}
\usepackage{siunitx}
\usepackage[version=4]{mhchem}
\usepackage{graphicx}
\usepackage{caption}
\usepackage{subcaption}
\usepackage{cite}
\usepackage{geometry}
\usepackage[skip=2pt,font=footnotesize,margin=1cm]{caption}

\usepackage{siunitx} 
\begin{document}
\begin{center}
\textbf{\Large Shot-noise-limited emission from interband and quantum cascade lasers}\\
\vspace{5mm}
Tecla Gabbrielli,\textsuperscript{1,2,*,\dag} Jacopo Pelini,\textsuperscript{2,3, \dag} Georg Marschick,\textsuperscript{4} Luigi Consolino,\textsuperscript{1,2} Irene La Penna,\textsuperscript{1,2} Jérôme Faist,\textsuperscript{5} Mathieu Bertrand,\textsuperscript{5} Filippos Kapsalidis,\textsuperscript{5} Robert Weih,\textsuperscript{6} Sven Höfling,\textsuperscript{7} Naota Akikusa,\textsuperscript{8} Borislav Hinkov\textsuperscript{4,9,*}, Paolo De Natale,\textsuperscript{1,2} Francesco Cappelli,\textsuperscript{1,2,$\ddagger$} and Simone Borri\textsuperscript{1,2,$\ddagger$}
\\
\bigskip
\textit{\small\textsuperscript{1} CNR-INO -- Istituto Nazionale di Ottica, Via Carrara, 1 - 50019 Sesto Fiorentino FI, Italy}
\\
\vspace{1mm}
\textit{\small\textsuperscript{2}LENS -- European Laboratory for Non-Linear Spectroscopy, Via Carrara, 1 -- 50019 Sesto Fiorentino FI, Italy}
\\
\vspace{1mm}
\textit{\small\textsuperscript{3}University of Naples Federico II, Corso Umberto I 40 -- 80138 Napoli, Italy}
\\
\vspace{1mm}
\textit{\small\textsuperscript{4}TU Wien -- Institute of Solid State Electronics \& Center for Micro- and Nanostructures, Gußhausstraße 25-25a -- 1040 Vienna, Austria.}
\\
\vspace{1mm}
\textit{\small\textsuperscript{5}Institute for Quantum Electronics, ETH Z\"{u}rich, Z\"{u}rich, Switzerland}
\\
\vspace{1mm}

\textit{\small\textsuperscript{6}Nanoplus GmbH, Oberer Kirschberg 4, 97218 Gerbrunn, Germany}
\vspace{1mm}
\textit{\small\textsuperscript{7}Julius-Maximilians-Universität Würzburg, Physikalisches Institut, Lehrstuhl für Technische Physik, Am Hubland, 97074 Würzburg, Germany }
\\
\vspace{1mm}
\textit{\small\textsuperscript{8} SSD advanced innovation headquarters, Hamamatsu Photonics K.K., Shizuoka 431-2103, Japan }
\\
\vspace{1mm}
\textit{\small\textsuperscript{9} Silicon Austria Labs, Europastraße 12, 9524 Villach, Austria} \\
\textit{\small\textsuperscript{$\dag$}These authors equally contributed to this work.} \\
\textit{\small\textsuperscript{$\ddagger$} These authors jointly supervised this work.}
\\
\bigskip
* tecla.gabbrielli@ino.cnr.it, borislav.hinkov@silicon-austria.com
\end{center}



\begin{abstract} 
The intensity noise of a laser source represents one of the key factors limiting the ultimate sensitivity in laser-based systems for sensing and telecommunication. For advanced applications based on interferometry, the availability of a shot-noise-limited local oscillator is even more important for the effective feasibility of high-precision measurements. This is particularly crucial in quantum optics applications based on homodyne detection schemes to measure non-classical light states, such as squeezed states. This work deeply investigates and analyzes the intensity noise features of the most widely used mid-infrared semiconductor heterostructured lasers: quantum cascade and interband cascade lasers. For this purpose, a comprehensive comparison of three different continuous-wave lasers operating at room temperature around \SI{4.5}{\micro m} wavelength is presented. First, a thorough electro-optical characterization is given, highlighting the differences and the shared common characteristics of the tested devices. Then, a detailed intensity noise analysis is reported, identifying their different noise operations with a particular reference to shot-noise-limited operations. Finally, some perspectives towards advanced applications are discussed.

\end{abstract}

\section{Introduction}
The availability of high-performance semiconductor lasers capable of generating mid-infrared (MIR) coherent light has provided a determinant boost to the development of MIR optical technologies, enabling a large variety of applications. Typical examples include molecular spectroscopy and trace-gas sensing, which are pivotal for the present and future challenges in environmental sustainability and security applications. These applications benefit from the linestrength of the fundamental molecular absorption lines in the MIR and would significantly benefit from higher detection sensitivity. Scientific and technological outcomes in these research fields impact food analysis~\cite{Szwarcman2021,Dabrowska2022}, environmental control~\cite{Galli:2011,DelliSanti:2022}, medical diagnostics~\cite{PleitezRafael2013,Kumar2018}, pharmaceutical analysis~\cite{Hinkov2023,Altug2022,Pilat2023}, and security testing~\cite{Hinkov2010}. Moreover, the presence of atmospheric high-transparency windows makes the MIR very attractive also for free-space communication applications~\cite{Seminara:22,Su:18, Flannigan2022}, driving further technological developments such as IR-Radars~\cite{hanson2001single} and telecommunication systems~\cite{Pang2022_6gbps,Dely2022, Corrias:2022}. Motivated by these numerous applications, the scientific community and industry are making an intense effort towards the development of sensing technologies being able to significantly improve relevant figures-of-merit to unprecedented levels, including selectivity and sensitivity~\cite{Wang:2022,Jiang:2023}.
A deeper understanding of the noise characteristics of the laser sources is at the core of such further advancements. Depending on the application, the intensity noise of the used laser source can represent one of the key parameters defining the achievable detection limit (sensitivity) of a sensor~\cite{lawrie2019quantum}. Using devices capable of emitting shot-noise-limited (or even sub-shot-noise) radiation can potentially lead to the development of imaging and spectroscopic systems with significantly improved resolution and sensitivity~\cite{Hodgkinson2013,frascella2021overcoming} or to communication setups with suppressed bit-error rate data transmission~\cite{Tyson2002}. Shot-noise-limited operation is also crucial for high-precision interferometry \cite{mcculler2020frequency}, and can be considered as a prerequisite for the future development of quantum-optics-based homodyne detection schemes, where shot-noise-limited local oscillators are relevant for measuring non-classical light state emission, e.g. squeezed states~\cite{Gabbrielli:2021}.
\\ 
To date, the most widespread mid-infrared laser sources are Quantum Cascade Lasers (QCLs)~\cite{Faist:1994} and Interband Cascade Lasers (ICLs)~\cite{Yang1995}. Both are based on complex semiconductor heterostructure active regions and are capable of high-performance room-temperature pulsed or continuous-wave (CW) operation~\cite{silva2005integrated,yu2006temperature}. They are characterized by a periodically cascading structure, resulting in an enhancement of the emitted optical power. Importantly, if taken together, ICLs and QCLs cover almost the entire MIR spectral range above $\sim$\SI{2.6}{\micro m} wavelength up to the THz range~\cite{Kohler:2002, Meyer:2020}. Moreover, the range between roughly 4.0 and \SI{6.0}{\micro \meter} is covered by both types of sources. Given the importance of these two sources for the MIR region, a comparative analysis of their intensity noise properties is of particular interest, playing a crucial role in selecting the optimal source for specific next-generation applications. The comparison is even more relevant due to their fundamentally different internal carrier dynamics, despite their aforementioned similarities~\cite{Faist:2013bo,Trombettoni:2021,Meyer:2020}.

QCLs are unipolar lasers~\cite{Faist:2013bo}, where the fast laser transition ($\tau < \SI{1}{ps}$) occurs between intersubband levels in the conduction band. Due to this fast dynamics, QCLs are characterized by relatively high threshold current densities (typically in the kA/cm$^2$ range~\cite{Hinkov2012}). In return, they benefit from a particularly narrow Schawlow-Townes linewidth of $\approx \SI{100}{Hz} $~\cite{Bartalini:09}, which makes them appealing for high-resolution spectroscopy and frequency metrology applications~\cite{Galli:2013b,Consolino:2018,Borri:2019a}. Additionally, they can be modulated at GHz frequencies~\cite{Paiella2000,Hinkov2016}, a key feature for free-space communication applications. In the last decade, comb emission from broadband multimodal QCLs~\cite{Riedi:2015} ("QCL-combs") was demonstrated~\cite{Hugi:2012}, based on the third-order nonlinearity of their active medium~\cite{Friedli:2013,Faist:2016}. The consequential development of tailored QCL-combs has revolutionized the research field of MIR precision measurements, providing controllable direct frequency comb emitters at these wavelengths~\cite{Burghoff:2014,Burghoff:2015,Singleton:2018,Cappelli:2015,Cappelli:2016,CappelliConsolino:2019,Consolino:2021b, Eramo24}. They could be successfully exploited for dual-comb spectroscopy~\cite{Villares:2015a} and free-space optical communication~\cite{Corrias:2022}.

ICLs are bipolar devices and, similar to standard diode lasers, their lasing process is based on interband transitions between the conduction and valence band, resulting in a much longer upper carrier-level lifetime ($\approx \SI{1}{ns}$) than QCLs~\cite{vurgaftman2015interband}. Because of this latter feature, ICLs can operate at a much lower bias current density (typically $\ll$~kA/cm$^2$~\cite{Vurgaftman2013,Knotig2020}) as compared to QCLs~\cite{Hinkov2012}. This is a significant advantage in energy consumption and heat dissipation. These features are especially relevant in compact outdoor in-field and remote applications where the supplied energy resources are strongly limited compared to in-lab applications. The nature of their interband transition hampers ICLs from operating in the long-wavelength infrared range (current wavelength limit $\simeq$~\SI{13}{\micro m} \cite{Massengale:22}). In return, they can be easily designed to operate in the 3-\SI{4}{\micro m} range, where QCLs show limited performance. This makes ICLs highly suitable for particular spectroscopic/sensing applications, e.g. measurements of the C-H fundamental band falling in this spectral range \cite{Ye:16,Xia:2021}, and for free-space communication in the 3-\SI{4}{\micro m}-transparency-window of the atmosphere~\cite{Didier:22,spitz2022interband}.
Similar to QCLs, ICLs can work both in pulsed and CW operation~\cite{kim:2008,Canedy:14}. However, the emitted optical power is typically lower, due to a lower number of active region stages (typically 3-10 periods) than in QCLs (typically tens of periods). Despite the lower number of periods, CW ICLs with 10-stages active medium were demonstrated to emit up to \SI{500}{mW} of output power at room temperature~\cite{Meyer:2020}, at the cost of a higher threshold current comparable to low-dissipation QCLs \cite{Bismuto:2015}. Recently, the possibility of comb emission with ICLs was demonstrated. As in the case of QCLs, the comb emission is enabled by the third-order nonlinearity of their active medium~\cite{Meyer:2020}. This discovery opens the doors to ICLs' exploitation as sources for advanced precision and metrological applications \cite{Marschick:2024}. \\Triggered by potential applications in metrology, the frequency noise characteristics of both QCLs and ICLs have been investigated in several works \cite{Bartalini:2010,Bartalini:2011,Borri:2020}. In contrast, at present, only a few studies focus on the investigation of their intensity noise features \cite{Gensty:2005, Simos:14,Juretzka:15}. According to measurements reported in literature, the Intensity Noise Power Spectral Density (INSPD) of QCLs is characterized by a 1/f trend at low frequencies (typically lower than a few MHz), due to electron tunneling dynamics through the multi-barrier structure~\cite{Gabbrielli:2022,Borri:2011}. In ICLs,
a $1/f^2$ trend at low frequencies (approximately within the range 5 kHz - 20 kHz) has been observed at low temperatures~\cite{folkes2008interband}. 

In this work, we thoroughly analyze and compare the Intensity Noise Power Spectral Density (INPSD) of three CW-emitting devices with similar emission wavelengths of around \SI{4.5}{\micro m} at room temperature: a commercial Distributed-Feedback (DFB) QCL characterized by a particularly low threshold current, a custom-made Fabry-Perot (FP) QCL with multiple (single-mode and comb) operating regimes, and a commercial DFB ICL. The lasers are first compared in terms of their main features, light-current-voltage (LIV) characteristics, and emission spectra. Their INPSD is then characterized via a custom-designed high-bandwidth MIR balanced detector \cite{Gabbrielli:2021}: the acquired signals enable the study of the intensity noise of the collected radiation in a broad Fourier frequency range (1-\SI{100}{MHz}) while simultaneously monitoring the corresponding shot-noise level and, finally, highlighting shot-noise limited operation regimes. This study opens the doors to practical applications of MIR shot-noise-limited local oscillators in balanced homodyne detectors \cite{Gabbrielli:2021} and other interferometric measurement techniques, where the light source noise is determinant for the ultimate precision limit \cite{lawrie2019quantum,PhysRevLett.123.231107,mcculler2020frequency}.
\section{Methods and Discussion}

\subsection{Main features of the lasers}
\begin{figure}[!htb]
    \centering
    \includegraphics[width = 1.05
\textwidth]{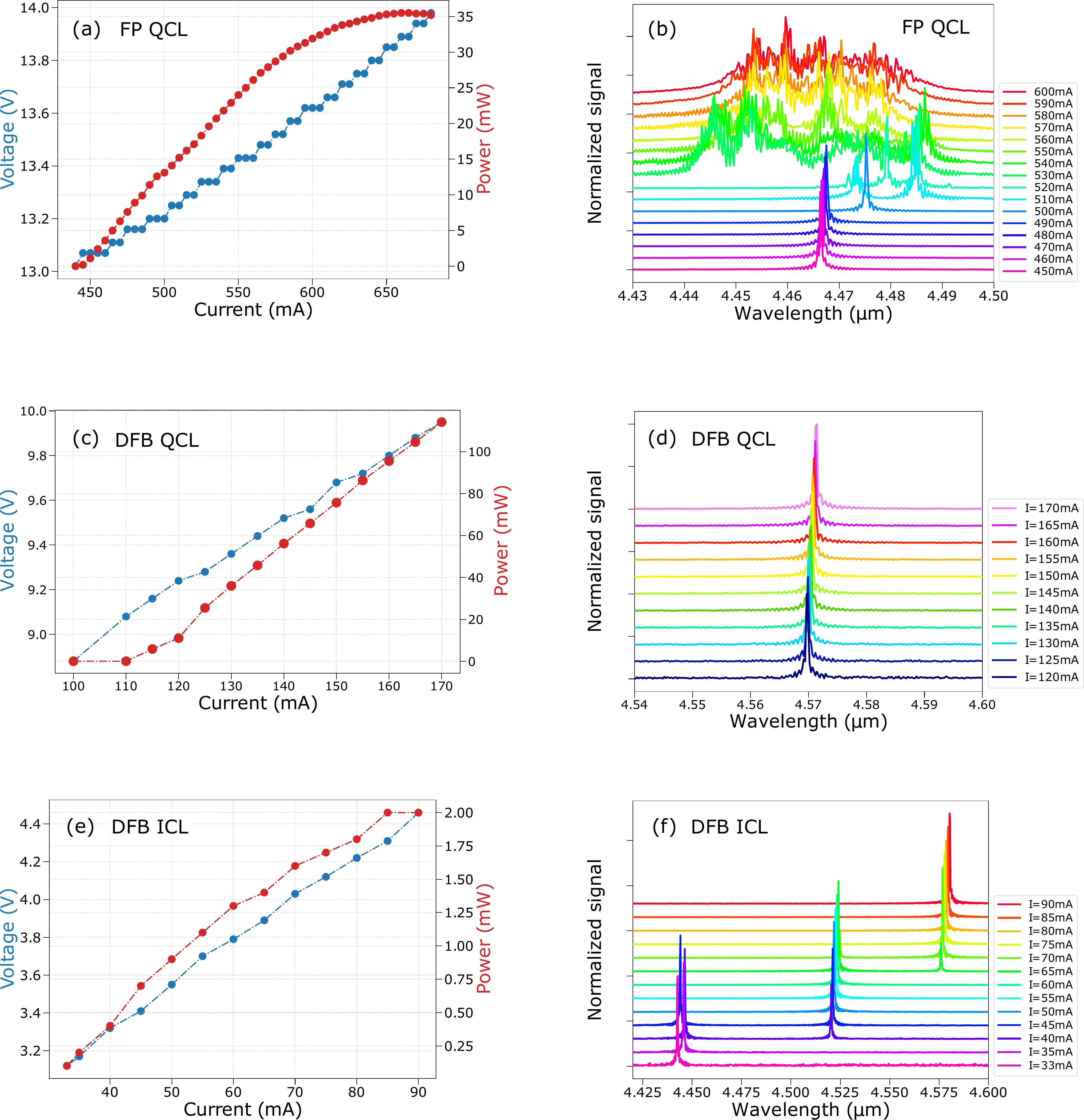}
    \caption{LIV curves of (a) the FP QCL, (c) the DFB QCL, and (e) the DFB ICL,  and the corresponding emission spectra (b,d,f) measured at a fixed temperature of \SI{20}{\celsius}. The emitted power is depicted in red, and the voltage data is shown in blue. The emission spectra (b,d,f) are acquired via an Optical Spectrum Analyzer (OSA - FTIR, Bristol) for several values of the laser bias current I, with the values reported in the legend. Each emission spectrum is normalized to its maximum peak intensity.}
    \label{fig:QCL_LIV}
\end{figure}
The three devices analyzed in this work are representatives of the main typologies of current state-of-the-art MIR lasers: two single-mode devices (the DFB QCL and the DFB ICL) and a comb-emitting laser (the FP QCL).

The DFB ICL is a typical ICL designed for spectroscopy and sensing applications. It is characterized by low power consumption and a wide tuning range, making it particularly suitable in all applications requiring low energy consumption, like outdoor in-field sensing.

The DFB QCL is the most commonly used type of MIR laser in current laboratory setups and commercial sensing applications. The selected device belongs to the latest generation of QCLs and it is provided with some appealing features like the butterfly mount and the engaging low threshold current. It can be considered a very good compromise between energy consumption and optical emission power, with the additional advantage of mode-hop-free tuning in the tested bias current range. 

Finally, the FP QCL is a widely tunable and versatile device, whose most compelling feature is represented by the different emission regimes (single mode, bi-lobed comb, harmonic comb, and dense comb), which can be accessed by simply tuning the driving current while keeping the other parameters constant. This makes such devices suitable for a variety of scenarios, including challenging sensing applications, metrology, dual-comb sensing, free-space data communication, and many more. 

In order to select the proper operating conditions for the intensity noise analysis, the key properties of the three devices, including the LIV (Light-Current-Voltage) curves and their emission spectra (Fig.~\ref{fig:QCL_LIV}) have been first characterized. The characterization described in this work is performed at room temperature (\SI{20}{\celsius}), except when explicitly mentioned. The structural details of these devices are reported in Appendix Section \ref{appendix:devicestructure}.

The first characterized device is the FP QCL, from ETH Z\"urich, which is particularly interesting for its versatile modes of operation. By varying its operating conditions (e.g. driving current), it is possible to switch its emission regimes, as described below. The LIV curve of the device is shown in Fig.~\ref{fig:QCL_LIV}(a): the laser threshold is around \SI{450}{mA} when operated at room temperature (\SI{20}{\celsius}), and it reaches a CW optical emission power of \SI{35}{mW} before its rollover at $\sim$\SI{650}{mA}. The spectral analysis in Fig.~\ref{fig:QCL_LIV}(b) shows the different operating regimes of this device. At the laser threshold, it operates in a single-mode regime with a wavelength slightly below \SI{4.47}{\micro m}. It remains single mode up to $\sim$\SI{510}{mA}, where the emission becomes bicolor, characterized by two peaks centred at \SI{4.473}{\micro m} and \SI{4.485}{\micro m}, respectively.. Between \SI{520}{mA} and \SI{530}{mA}, we can notice a harmonic comb emission, i.e. an emission regime characterized by a low number of correlated modes \cite{Gabbrielli:2022, Picardo18}. For higher currents, the device operates as a standard dense QCL comb. This QCL was electrically supplied via a modular current driver and controller from ppqSense s.r.l (QubeCL10), which is optimized for ultra-low-noise operation with a nominal current noise density of \SI{300}{pA /\sqrt{Hz}}. \\
The second tested device is the DFB QCL, from Hamamatsu Photonics, which has a particularly short waveguide (only \SI{2}{mm}-long vs \SI{4.5}{mm} of the previously discussed FP QCL as reported in Appendix Section \ref{appendix:devicestructure}, enabling a low threshold current operation ($\sim$\SI{110}{mA}). Compared to the tested FP QCL, this device also shows a higher quantum efficiency, i.e. the efficiency in transforming the injected electrons into emitted photons. The DFB QCL's LIV curve is shown in Fig.~\ref{fig:QCL_LIV}(c). The laser achieves a CW optical output power well beyond \SI{100}{mW} for a bias current of \SI{170}{mA}. Another peculiarity of the laser is its butterfly mount, which makes the device particularly easy to handle. Due to current constraints and butterfly mount, this device was driven with a customized QubeDL made by ppqSense s.r.l, characterized by a nominal current density noise of $\SI{500}{pA/\sqrt{Hz}}$. In our experiment, the laser was directly integrated into one of the modules of the current driver, to minimize parasitic effects and interference from external electromagnetic noise sources. As shown in Fig. \ref{fig:QCL_LIV}(d), the laser operates at a wavelength around  \SI{4.57}{\micro m} and in a single-mode emission in the whole tested current range. \\ 
The third device is a ridge DFB ICL produced by nanoplus, driven via the same ultra-low-noise driver as the previously discussed FP QCL. Compared to the QCLs described above, this device has a lower number of active stages (6 vs 25 in the case of the DFB QCL) as described in detail in Appendix Section \ref{appendix:devicestructure}. This feature, together with the smaller device dimensions (waveguide's length of only \SI{0.9}{mm}), results in a lower optical emission power with a maximum value of \SI{2}{mW} at \SI{20}{\celsius} temperature and \SI{90}{mA} driving current (see Fig. \ref{fig:QCL_LIV}(e)). As expected, the ICL is characterized by a significantly lower energy consumption than the QCLs, with a threshold bias current of $\sim$\SI{33}{mA}.
The ICL emission spectra at room temperature are depicted in Fig. \ref{fig:QCL_LIV}(f) for different values of the bias current (I). Close to threshold, the laser emits in single mode at a wavelength around \SI{4.44}{\micro m}. For $\SI{40}{mA} < I < \SI{50}{mA}$, the ICL emits in a bicolor regime, with the modes centred at \SI{4.44}{\micro m} and \SI{4.525}{\micro m}, respectively. The laser alternates single mode and bicolor operation mode for higher currents, as shown in the figure.

In the following intensity noise analysis, all the devices have been operated in their single-mode emission ranges to compare them under similar operating conditions.

 \subsection{Intensity noise analysis}
As widely discussed in the introduction, a thorough comparison of the intensity noise behaviour of these devices is particularly interesting in identifying the most suitable source for each specific application. In particular, the search for laser sources characterized by shot-noise-limited or even sub-shot-noise emission is of great interest for the emerging technological era of ultra-precision and quantum measurements based on non-classical emission states~\cite{lawrie2019quantum}.

\begin{figure} [htb!]
    \centering
    \includegraphics[width=0.7\columnwidth]{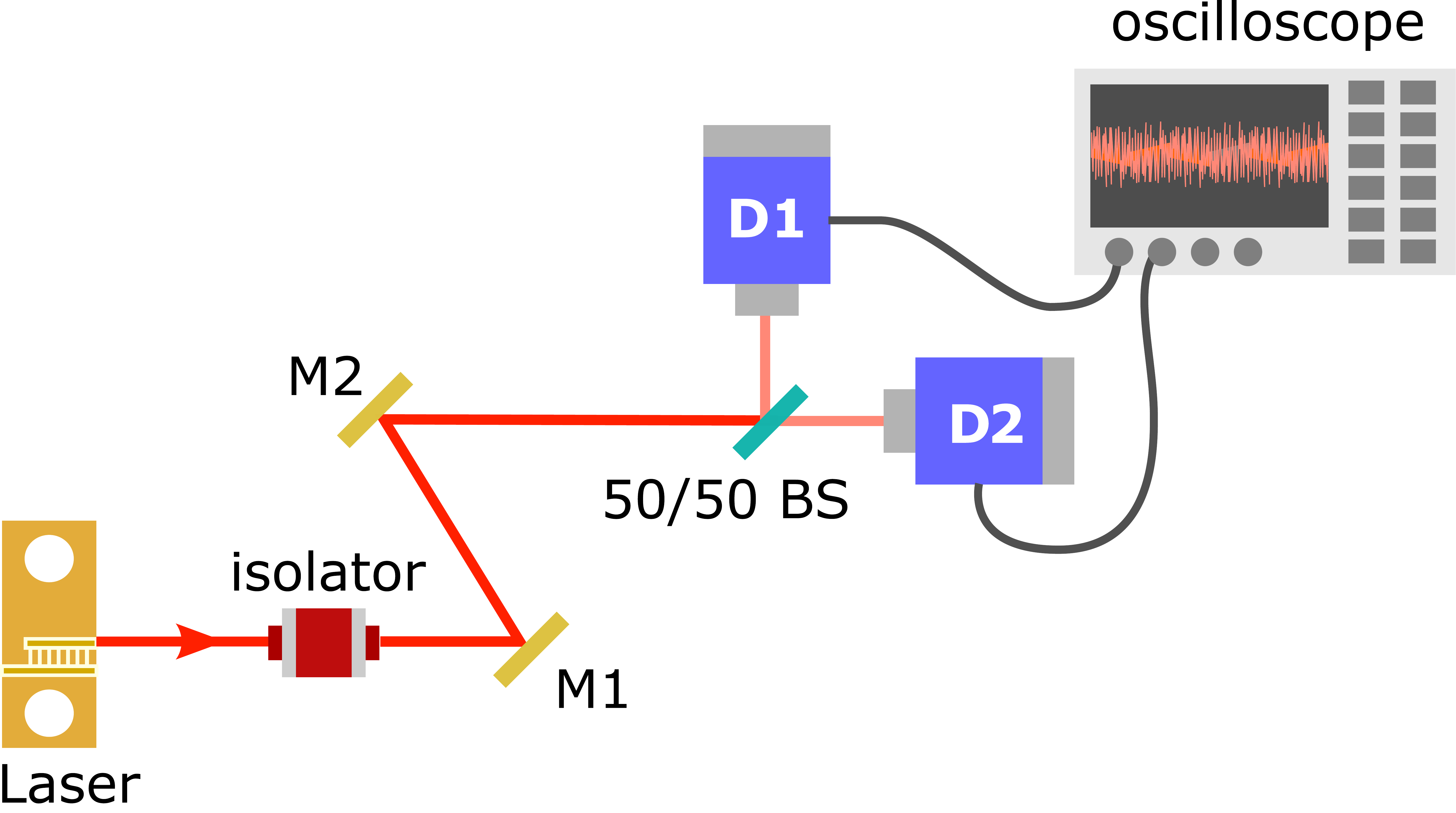}
    \caption{Sketch of the balanced detection setup used to characterize the laser sources. The optical isolator is used to avoid feedback from the optical elements into the laser source (the isolation level is 33 dB). M1 and M2 are two gold mirrors used for the optical alignment of the laser beam. 50/50 BS is a 50/50 beam splitter, and D1 and D2 are the two photovoltaic detectors. The photocurrents generated by D1 and D2 (mean value and amplified noise) are acquired in the time domain via an oscilloscope. 
    }
    \label{fig:setup}
\end{figure}
\begin{figure}[!htb]
    \centering
    \includegraphics[width =1.0\textwidth]{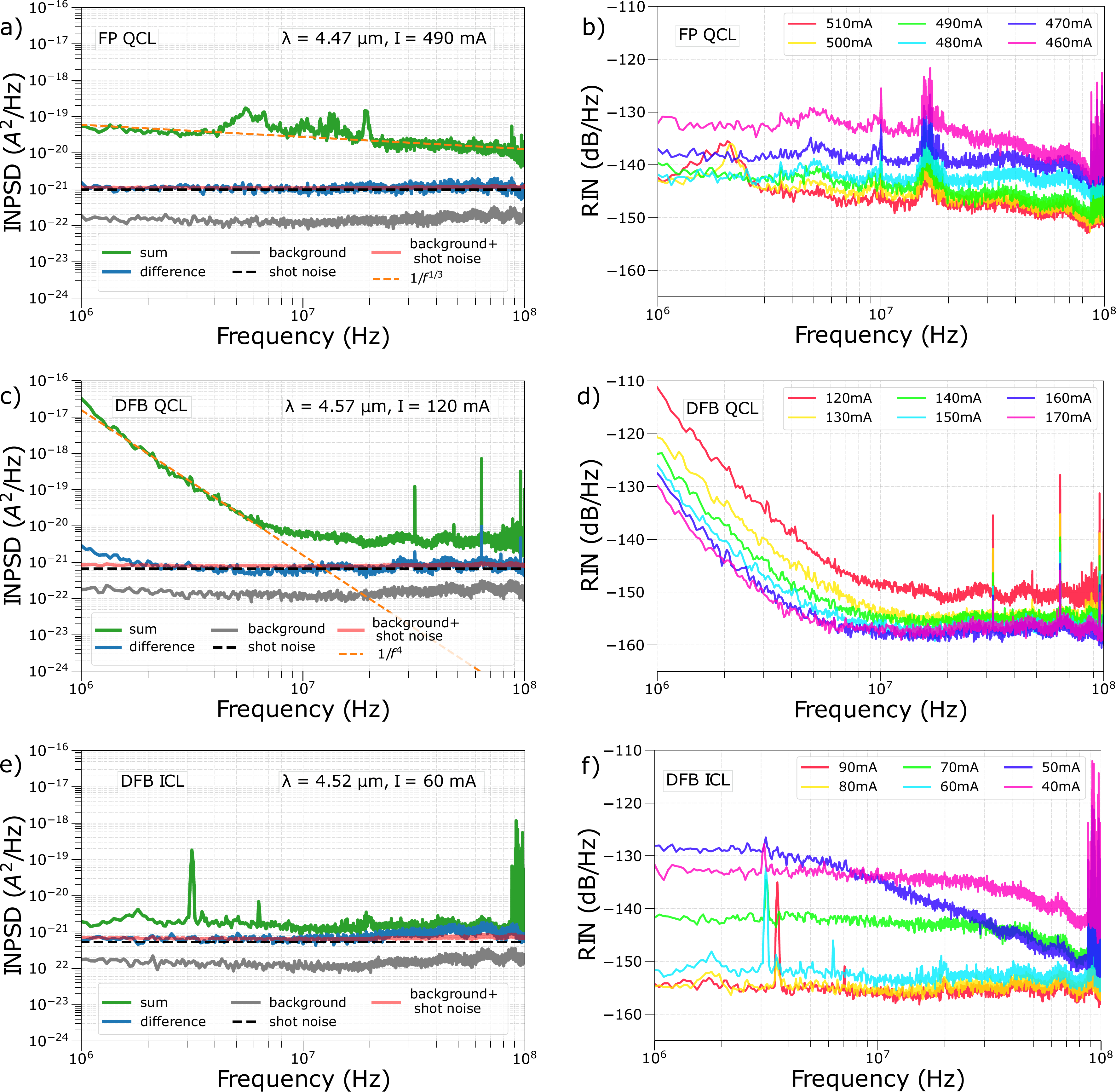}
    \caption{INPSDs of the (a) FP QCL, (c) DFB QCL and (e) DFB ICL operated at a fixed temperature of \SI{20}{\celsius} and at a bias current of \SI{490}{mA}, \SI{120}{mA}, and \SI{60}{mA}, respectively. The INPSDs of the sum (green traces) and the difference (blue traces) are compared to the expected shot-noise level (black dashed lines), to the INPSD of the sum of the background noise (grey traces), and the sum of these two latter quantities (i.e. background + shot noise, red traces). In the INPSDs of the sum of both the DFB QCL (c) and the ICL (e), there are visible peaks (above \SI{20}{MHz} and around \SI{3}{MHz}, respectively) due to technical noise, associated to resonances in the mass loop of the supply chain (i.e. voltage supply of the QubeCL, and the QubeCL itself). In the ICL INPSD of the sum at around \SI{100}{MHz} there are visible peaks due to pick-up noise from transmitted FM radio signals. In graphs (a) and (c) the $1/f^{1/3}$ and $1/f^4$ trend are plotted as orange dashed lines.  In graphs (b,d,f) the RIN of (b) the FP QCL, (d) the DFB QCL (d), and (f) the DFB ICL are reported for several values of the bias current, as reported in the legends. The RIN analysis is done at a fixed temperature of \SI{20}{\celsius}.} 
    \label{fig:INPSD}
\end{figure}
Measuring intensity noise down to the shot-noise level in the MIR range is a non-trivial task. First, it requires high-bandwidth detectors because the emission is usually dominated by $1/f$ or technical noise at low Fourier frequencies. Second, the dark noise (including electronic and background thermal noise contributions) has to lie sufficiently below the shot-noise level to be able to measure it properly. The measurements presented in this work are performed with a homemade balanced detector (Fig.~\ref{fig:setup}). This system has been designed and developed to fill the existing technological gap in the MIR~\cite{Gabbrielli:2021}. Based on a differential measurement at the output of two equal MCT photovoltaic detectors, its main advantage is a sufficiently large dynamical range to allow the suppression of common-mode noise in the two arms of a 50/50 beam splitter (BS) down to the corresponding shot-noise level. Moreover, the detection system (photodectors+oscilloscope) can rely on fast response, having an overall effective detection bandwidth of \SI{120}{MHz} ~\cite{Gabbrielli:2021}.
\\ The developed balanced detection system is here applied to measure the intensity noise of a MIR source working in the 4--5-\SI{}{\micro m} spectral range while monitoring the corresponding shot-noise level. 
In the performed experiment, each acquisition, lasting \SI{1}{ms}, consists of synchronously recording the output signals of the two MCT detectors via an oscilloscope.  
For each couple of acquired signals, the Fast Fourier Transform (FFT) of their sum and difference are then computed via a Python script, from which the corresponding INPSDs are obtained. The measurements are performed in the linear-responsivity regime of the detectors, avoiding any saturation. The detector responsivity, in the linear regime, is \SI{1.48}{A/W} at \SI{4.5}{\micro m}, leading to a quantum efficiency of up to 40\%~\cite{Gabbrielli:2021}. The saturation intensity of each detector at $\SI{4.5}{\micro m}$ is $\simeq$\SI{1}{mW}. Therefore, the incident power on the BS does not have to exceed $\simeq$\SI{2}{mW}, leading to a maximum achievable clearance (ratio between the shot-noise level and the background noise) of \SI{9}{dB}~\cite{Gabbrielli:2021}. Due to the saturation limit, when using laser sources as powerful as the QCLs presented in this work, their radiation impinging onto the BS needs to be attenuated accordingly.
\\
In Fig.~\ref{fig:INPSD}, the intensity noise analysis of the three devices is resumed.  Referring to the noise analysis, the INPSD of the sum (green traces in Fig.~\ref{fig:INPSD} (a), (c), and (e)) gives information regarding the intensity noise of the collected radiation, while the INPSD of the difference (blue traces in the figures) represents the corresponding shot-noise level.  
In Fig.~\ref{fig:INPSD} the expected shot-noise level is also shown (black dashed lines) for all the three lasers, calculated as $PSD_{\mathrm{SN}}= 2 e \left (I_{\mathrm{1}} + I_{\mathrm{2}}\right )$, where ($I_{\mathrm{1}}$ +$I_{\mathrm{2}}$) is the sum of the mean values of the currents measured at the two detectors' outputs and $e$ is the electron charge~\cite{Gabbrielli:2021}. 
These traces are also compared to the background noise (grey traces), which takes into account both the detector noise and the environmental thermal noise of the laboratory. Indeed, these traces are obtained by having the detector's switch on and measuring the detector's electronic background noise plus the room's thermal background noise while the laser is off.
Finally, the measured INPSDs are also compared to the sum of the expected shot-noise level and the background noise (red traces). As expected, the INSPD of the difference (e.g. the measured shot-noise level) overlaps with the red traces for all three lasers.

Concerning the comparison of the three lasers, some important results can be extrapolated from the graphs. It is worth noticing that the level of applied attenuation is a critical parameter in this kind of measurement. Indeed, attenuation alters, as a side effect, the statistics of the tested light towards a Poisson distribution. Consequently, an excess of attenuation could lead to a shot-noise-limited detection despite a laser emission with a noise level higher than the shot noise. In the case of the ICL, no attenuation was needed due to its low output power. Instead, in the case of the two QCLs, an attenuation was required to avoid detector saturation. Regarding the two QCLs, we decided to test the two devices in the same operation condition, i.e., reasonably above threshold and at relatively low power to limit the attenuation needed. The selected working conditions are $I=~\SI{490}{mA}$ for the FP QCL and $I=~\SI{120}{mA}$ for the DFB QCL, which correspond to a factor of 10\% above the laser threshold $I_{\mathrm{th}}$, i.e. $(I-I_{\mathrm{th}})/I_\mathrm{th} \simeq 0.10$. 
The attenuation level is 87~$\%$ in the case of the DFB QCL and 82~$\%$ in the case of the FP QCL. 
As depicted in Fig.~\ref{fig:INPSD}(a), the FP QCL shows a residual $1/f^{1/3}$ trend in its intensity noise (INPSD of the sum, Fig.~\ref{fig:INPSD}(a)). As a result, the INPSD of the sum lays well above the corresponding shot-noise level for the entire tested Fourier frequency range of 1--\SI{100}{MHz}. This residual excess noise decays smoother than the typical $1/f$ trend shown by such devices~\cite{Bartalini:2011,Yamanishi:2014,Yamanishi:2008}. 
\\ 
For the DFB QCL, a different current driver (customized QubeDL) had to be used, as already introduced in the previous section. In this case, an excess of noise on the measured intensity noise in the low Fourier-frequency range is present, as clearly visible in Fig.~\ref{fig:INPSD}(c) (sum signal, green trace). The steep decay of this excess noise follows a $1/f^4$ trend (dashed orange line Fig.~\ref{fig:INPSD}~(c)). Above \SI{10}{\mega \hertz} this excess noise becomes negligible, allowing a genuine analysis of the laser intensity noise. For this QCL, the comparison between the sum and the difference signals (green and blue traces in Fig.~\ref{fig:INPSD}(c), respectively) clearly shows that the measured intensity noise lies above the shot-noise level for the whole portion of the analyzed frequency range. 
\begin{figure}[!htb]
    \centering
   \includegraphics[width = 0.6\columnwidth]{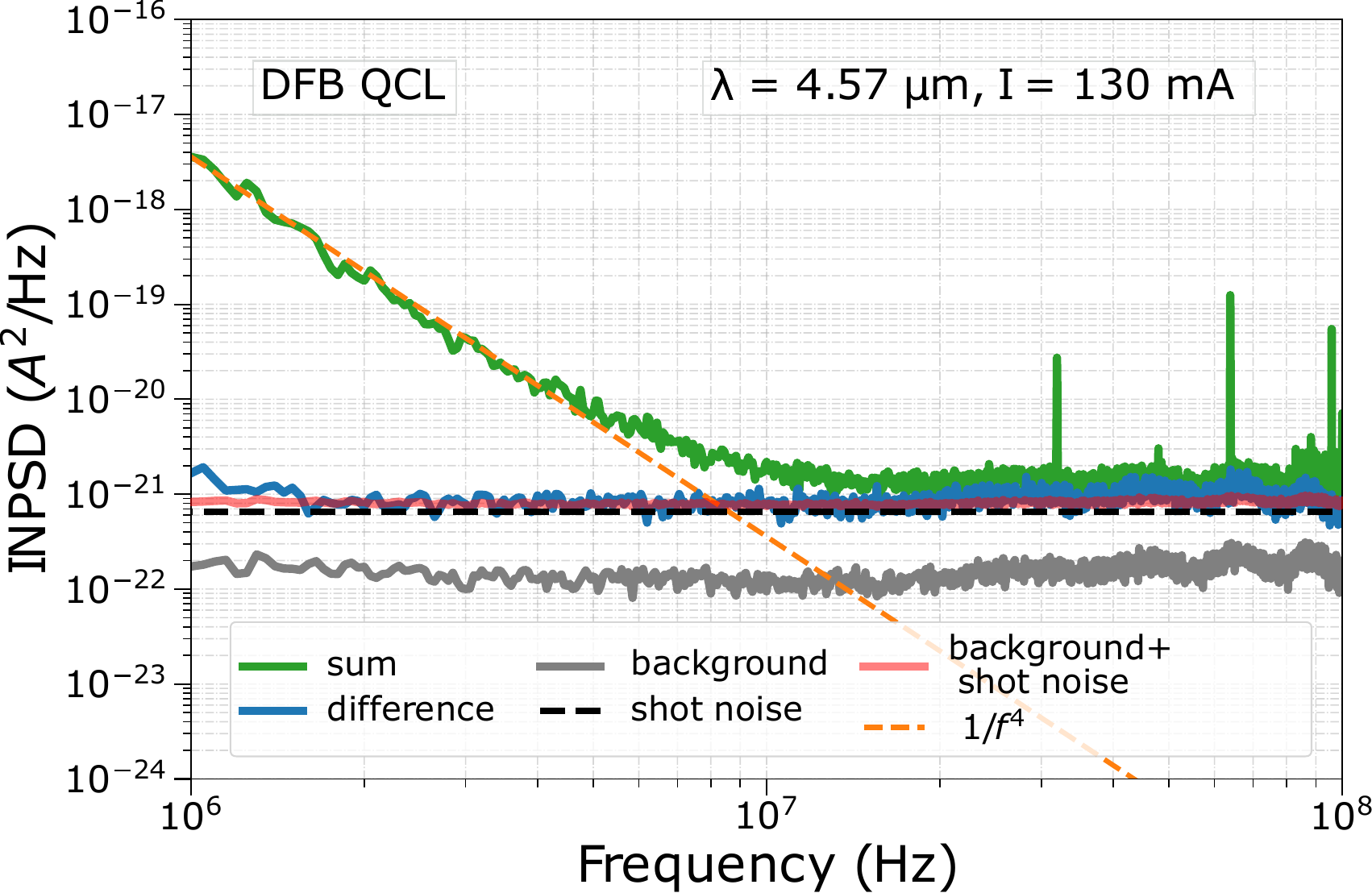}
 \caption{INPSD analysis performed for the DFB QCL operated at 130 mA and room temperature. The INPSD of the sum (green trace) and INPSD of the difference (blue trace) are compared to the expected shot-noise level (black dashed line), to the INPSD of the sum of the background noise (grey trace), and the sum of these two latter quantities (i.e. detector background + shot noise, red trace).}
  \label{fig:INPSD_DFBQCL}
\end{figure}
\begin{figure*}[!htb]
    \centering
    \includegraphics[width = 1.0\textwidth]{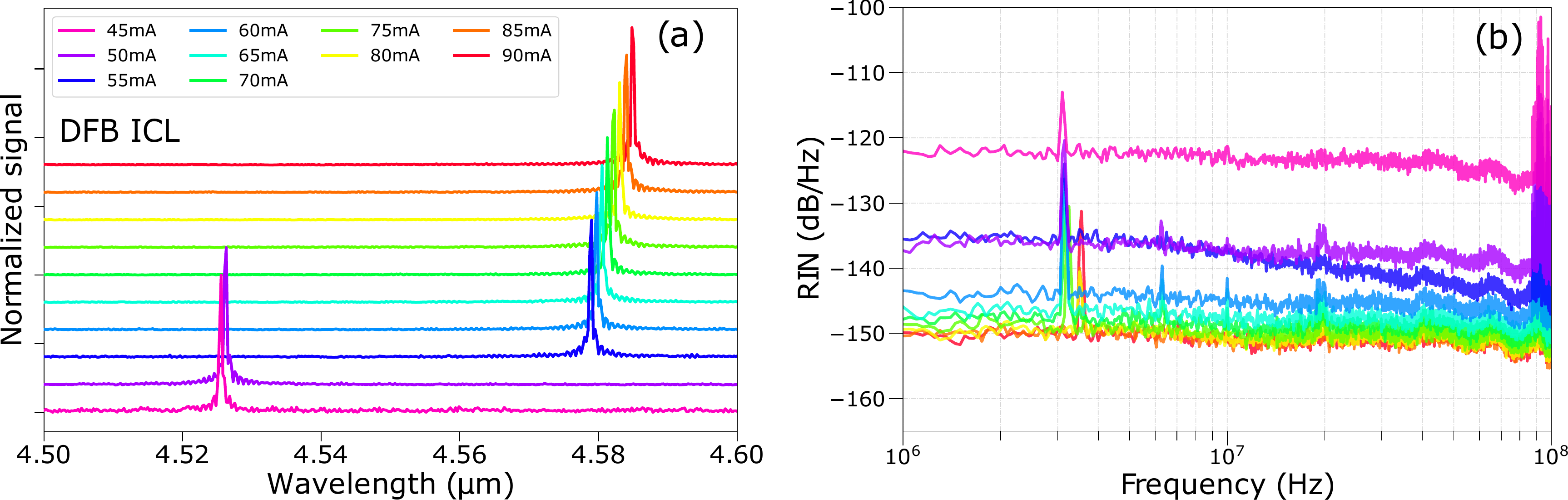}
    \caption{(a) Optical emission spectra and (b) RIN spectra of the ridge ICL performed at \SI{30}{\celsius} for several values of bias current. In this regime, the laser emission is always in a single mode.}
    \label{fig:RIN_30gradi}
\end{figure*}
\\
The DFB ICL is the device with the best performance in terms of low-intensity noise. Fig~\ref{fig:INPSD}(e) shows the INPSD traces of the ICL, operated in a single-mode emission regime. Its measured intensity noise is quite flat for the investigated Fourier-frequency range and tends to the white-noise behaviour expected by a perfect coherent source, i.e., dominated by Poisson statistics \cite{loudon:2000quantum}. In particular, the measured ICL radiation intensity noise flattens to the shot-noise level for Fourier frequencies above $\sim$\SI{10}{MHz}.
\\
As an additional analysis, in Fig. \ref{fig:INPSD_DFBQCL}, the INPSD of the DFB QCL is shown when operated at a higher current (130 mA). Here, a higher attenuation level (96\%) was required to avoid saturation of the detectors. It is interesting to note that, in this case, a shot-noise-limited radiation is achieved by considerably attenuating the laser emission. Indeed, the measured intensity noise (INPSD of the sum) is shot-noise-limited in the whole frequency range above \SI{10}{MHz}. This outcome, together with the shot-noise-limited results found for the DFB ICL, establishes the exploitability of both ICLs and QCLs to deliver shot-noise-limited radiation in the MIR range. Thus, this confirms their potential applicability in different practical scenarios contingent upon the laser source availability and desired wavelength specifications.

To complete the intensity noise analysis of the three devices, we also studied their Relative Intensity Noise (RIN) at different bias current values. The RIN is defined, in our case, as the INPSD of the sum normalized to the square of the total measured photo-current ($I_1 + I_2)^2$ following the same procedure used in Ref.~\cite{Gabbrielli:2022}. The results are reported in Fig.~\ref{fig:INPSD} (plots on the right column). Regarding the FP QCL, we have selected the bias current range where it operates in the single mode (up to \SI{500}{mA}, Fig.~\ref{fig:QCL_LIV}~(b)) and the bi-color (at \SI{510}{mA}) regime. The RIN is higher when the laser is close to its threshold (pink curve, Fig.~\ref{fig:INPSD}(b)). It starts to decrease with increasing current, reaching the lowest level for the tested currents of \SI{510}{mA}.
As expected, the RIN of the DFB QCL also exhibits a decreasing behaviour with increasing driving current (Fig.~\ref{fig:INPSD}~(d)). Similarly to the reported INPSDs of the sum, each single RIN trace follows a steep decreasing trend reaching almost a white-noise plateau for frequencies above \SI{10}{MHz}.
It is worth noting that the measured RIN levels are compatible with the values estimated by theoretical prediction~\cite{Simos:14}. In detail, a free-running QCL emitting 10--\SI{20}{mW} is expected to have a RIN of approximately $-(150-160)\SI{}{dB/Hz}$~\cite{Simos:14}.
Compared to existing literature, in particular, the previous experimental work~\cite{Juretzka:15}, our results show, for the first time, a good agreement with the expected theoretical behavior for both the QCLs, thus proving the theoretical model.  

Finally, the RIN values of the DFB ICL operated at~\SI{20}{\celsius} and different current values are reported in Fig.~\ref{fig:INPSD}~(f). These RIN traces correspond to the modal-jump behavior affecting the laser emission shown in Fig.~\ref{fig:QCL_LIV}~(e). The expected decreasing trend of the RIN for increasing driving current is broken in two cases due to an unstable regime connected with the bimodal emission. In particular, the graph shows a rise of the RIN by moving from the laser threshold ($I=\SI{40}{mA}$) to \SI{50}{mA}, where a first mode jump occurs. In the case of I = \SI{50}{mA}, a frequency cutoff at around \SI{4}{MHz} is present. This cutoff is well below the one due to the detector bandwidth and it is most probably due to the saturation effect driven by the high noise regime connected to the presence of unstable bimodal emission. An excess of noise is also present by moving the ICL current from \SI{60}{mA} to \SI{70}{mA}, corresponding to a second mode jump. In support of this hypothesis, unstable bimodal or few-mode emission regimes have already been connected to excess intensity noise, as shown in refs. \cite{Gabbrielli:2022, Chomet:2023}.
As evidence of the fact that this high noise regime is connected to the presence of the two-mode emission, and to complete the noise analysis for the ICL, Fig.~\ref{fig:RIN_30gradi} shows the ICL RIN measurements at a higher temperature (\SI{30}{\celsius}). As shown in Fig.~\ref{fig:RIN_30gradi}~(a), the laser emission remains single-mode for the entire tested range, even though a modal jump is present when the bias current is increased from \SI{50}{mA} to \SI{55}{mA}. The corresponding RIN curves are reported in Fig.~\ref{fig:RIN_30gradi}~(b). Apart from the modal jump between \SI{50}{mA} and \SI{55}{mA}, the RIN shows the expected clear monotonous decreasing trend with the bias current. \\ 
In addition, depending on the bias current, the measured ICL RIN levels reach values below -~\SI{150}{dB/Hz}, e.g. at room temperature for bias currents of \SI{60}{mA}, \SI{80}{mA} and \SI{90}{mA} (see Fig.~\ref{fig:INPSD}~(f)) or for currents above \SI{70}{mA} in the case of higher temperature (Fig.~\ref{fig:RIN_30gradi}~(b)). These results are compatible with those reported in literature for a ring ICL \cite{Marschick:2024}. Moreover, the DFB ICL tested in this work shows better performance with respect to previous works \cite{Didier:22,Deng2019-pk}, where RIN levels of -110~dB/Hz and -130~dB/Hz were reported, respectively. 

\section{Conclusion}
In this work, a thorough analysis of the intensity noise of the most popular MIR sources, ICLs and QCLs, is reported. Three devices have been investigated as representatives of the main typologies of state-of-the-art devices, namely a DFB ICL working in single-mode and bi-modal emission regimes, a single-mode DFB QCL, and an FP QCL working in both single-mode and comb regime, all operating at room temperature and characterized by the same emission wavelength of \SI{4.5}{\micro m}. Thanks to a custom-made MIR balanced detector with a 120-MHz bandwidth, the intensity noise features of the three devices have been analyzed over a wide Fourier frequency range, allowing us to identify the different noise contributions and unveiling, for the first time, their shot-noise-limited operation. 
In particular, we show that the measured ICL radiation INPSD lies close to the shot-noise level in the whole tested frequency range and substantially achieves the shot-noise for high Fourier frequencies. A higher noise level characterizes the radiation of the two QCLs, which have been operated at comparable working conditions in terms of relative distance from the threshold and output power. For these devices, the measured INPSDs of the sum remain above the corresponding shot-noise level for the entire tested frequency range, thus marking a difference with respect to the interband device. Nonetheless, we demonstrate the possibility of retrieving shot-noise-limited radiation starting from the emission of the DFB QCL when operated far from threshold at the cost of a sufficiently large attenuation.
Finally, a comprehensive RIN analysis was performed on all the tested devices. When the lasers are operated in single-mode regime, the RIN curves exhibit the expected decreasing trend with increasing laser bias current. This behavior is not always observed when the laser operates in multimodal emission regimes.
\\
The results demonstrated in this work, showing the availability of MIR shot-noise-limited radiation from ICLs and, under certain conditions, from QCLs, pave the way to improvements in the measurement sensitivity whenever the intensity noise of the source represents the limiting factor. Moreover, the availability of shot-noise-limited emitters can also allow the measurement, for the first time, of quantum correlations, that are expected in MIR cascade lasers due to the presence of FWM \cite{Gabbrielli:2022}. 
In this context, the experimental results shown here for the first time can provide useful insights for the development of new-generation MIR emitters capable of achieving low-noise performance undisclosed so far.

\appendix
\section{Devices' structure}
\label{appendix:devicestructure}

\textbf{Structural details of the FP QCL.}
The FP QCL has been fabricated at ETH Z\"urich. The active medium was grown via molecular-beam epitaxy and comprises two gain stages composed of alternate layers of \ce{In_{0.684}Ga_{0.316}As}/\ce{In_{0.335}Al_{0.665}As}, resulting in 2 gain stages (one centered at $\SI{4.6}{\micro m}$, the other one at \SI{4.3}{\micro m}). The active region (AR1+AR2) has a total thickness of \SI{2.1385}{\micro m}, with 17 periods in AR1 and 18 in AR2. The laser waveguide is \SI{4.5}{mm} long and has one high-reflectivity (HR) coated facet (reflection of approximately 100\%), while the other facet is uncoated and therefore characterized by a residual reflection of 27\%.
\vspace*{1mm}

\textbf{Structural details of the DFB QCL.}
The DFB QCL is a commercial device from Hamamatsu with a built-in DFB grating for single-mode emission, which is mounted on a butterfly housing. It is based on an InGaAs/InAlAs active medium with 25 stages and grown on an InP substrate. The active core is \SI{1.325}{\micro m} thick, and the laser waveguide is fabricated in an InP-based planar buried heterostructure geometry with a length of \SI{2}{mm}. Its uncoated outcoupling facet is characterized by a reflectivity of around 30~\%, while the back facet has an HR coating, reaching a reflectivity of above 98\%.

\vspace*{1mm}

\textbf{Structural details of the DFB ICL.}
The ridge DFB ICL is produced by nanoplus. This device has a cavity length of \SI{900}{\micro m} with a ridge width of \SI{7.35}{\micro m} and a 6-stage active medium enclosed by \SI{400}{nm}-thick \ce{GaSb} separate confinement layers and \ce{InAs}/{AlSb} superlattice cladding layers, which have a thickness of 3.5 and \SI{2.0}{\micro m}, respectively. A W-quantum well sequence is chosen to achieve a centre gain wavelength of \SI{4.5}{\micro m} (\SI{2.50}{nm} \ce{AlSb} /\SI{2.22}{nm} \ce{InAs} /\SI{2.50}{nm} \ce{In_{0.35}Ga_{0.65}Sb}/\SI{1.83}{nm} \ce{InAs}/\SI{1.0}{nm} \ce{AlSb}). The laser has 100 \% coating in the back facet and the front facet is as cleaved with a 30 \% reflectivity.

\section*{Funding}
The authors acknowledge financial support by the European Union’s NextGenerationEU Programme with the I-PHOQS Infrastructure [IR0000016, ID D2B8D520, CUP B53C22001750006] "Integrated infrastructure initiative in Photonic and Quantum Sciences'', by the European Union’s Research and Innovation Programmes Horizon 2020 and Horizon Europe with the Qombs Project [G.A. n.~820419] "Quantum simulation and entanglement engineering in quantum cascade laser frequency combs", the Laserlab-Europe Project [G.A.~n.~871124], and the MUQUABIS Project [G.A.~n.~101070546] "Multiscale quantum bio-imaging and spectroscopy", by the European Union’s QuantERA II [G.A.~n.~101017733] -- QATACOMB Project "Quantum correlations in terahertz QCL combs'', by the Italian ESFRI Roadmap (Extreme Light Infrastructure -- ELI Project), and by the European Union’s EDF Projects ADEQUADE "Advanced, Disruptive and Emerging QUAntum technologies for DEfense" (ID
101103417). \\ The authors are grateful for financial support from the Austrian Research Promotion Agency (FFG) through the ATMO-SENSE project [G.A.~n.~1516332] "Novel portable, ultra-sensitive, fast and rugged trace gas sensor for atmospheric research based on photothermal interferometry" and the NanoWaterSense project [G.A.~n.~873057] "Mid-IR Sensor for Trace Water Detection in Organic Solvents, Oils and Petrochemical Products". 
\section*{Acknowledgment}
The authors acknowledge the deep exchanges of ideas and fruitful conversations with the CNR-INO researchers Davide Mazzotti, Iacopo Galli, and Pablo Cancio Pastor concerning the noise trend evidenced in the here-reported analysis. The authors gratefully thank the company ppqSense s.r.l. for providing CNR-INO with the lasers' low-noise power supplies (QubeCLs). 
\section*{Disclosure}
The authors declare no conflicts of interest.
\section*{Data Availability}
The data supporting this study’s findings are available from the corresponding author upon reasonable request.

\bibliographystyle{ieeetr_CA}

\bibliography{sample}

\end{document}